\documentclass[prb,showpacs]{revtex4}

\usepackage{amsmath}

\begin{document}

\title{Exact and approximate expressions for the period of anharmonic oscillators}

\author{Paolo Amore}

\email{paolo@ucol.mx}

\affiliation{Facultad de Ciencias, Universidad de Colima,\\
Bernal D\'{i}az del Castillo 340, Colima, Colima,\\
Mexico.}

\author{Francisco M. Fern\'{a}ndez}

\email{fernande@quimica.unlp.edu.ar}

\affiliation{INIFTA (Conicet, UNLP), Blvd. 113 y 64 S/N, Sucursal
4, Casilla de Correo 16, \\ 1900 La Plata, Argentina}

\begin{abstract}
In this paper we present a straightforward systematic method for the exact
and approximate calculation of integrals that appear in formulas for the
period of anharmonic oscillators and other problems of interest in classical
mechanics.
\end{abstract}


\maketitle

\section{Introduction}

The discussion of periodic motion in one dimension is important in most
introductory courses on classical mechanics. Several problems can be solved
exactly, but in most cases one has to resort to approximate solutions.
Simple but sufficiently accurate approximate solutions for such problems are
very important in understanding many features of classical mechanics. In
addition to it, in some cases one is simply satisfied with accurate
numerical results, and expressions suitable for computation are most welcome.

The purpose of this paper is the discussion of the exact and approximate
calculation of the period of a particle that moves in one dimension under
the effect of an anharmonic potential.

\section{Periodic motion in one dimension}

Consider a particle of mass $m$ moving in one dimension under a
potential--energy function $V(x)$. Without loss of generality we assume that
$V(x)$ has a minimum at $x=0$; more precisely, we assume that $V(0)=0$, $%
V^{\prime }(0)=0$, and $V^{\prime \prime }(0)>0$, where the prime indicates
differentiation with respect to $x$. Following the standard notation in
classical mechanics, we use a dot to indicate differentiation with respect
to time, for example: $v=\dot{x}$.

>From the equation of motion
\begin{equation}
m\ddot{x}=-V^{\prime }(x)  \label{eq:Newton}
\end{equation}
we easily obtain an integral of the motion
\begin{equation}
E=\frac{m\dot{x}^{2}}{2}+V(x)  \label{eq:energy}
\end{equation}
which is the total energy. The motion of the particle is restricted to the
interval $x_{-}<x<x_{+}$, where the turning points $x_{\pm }$ satisfy $%
V(x_{\pm })=E$; that is to say, $\dot{x}=0$ at those points.

It is well--known that the period of the motion is given by
\begin{equation}
T=\oint dt=\sqrt{2m}\int_{x_{-}}^{x_{+}}\frac{dx}{\sqrt{E-V(x)}}
\label{eq:T_V}
\end{equation}
from which we obtain the frequency $\Omega =2\pi /T$.

We can simplify the equations of motion by the introduction of a
dimensionless time $\tau =\omega _{0}t$, where $\omega _{0}$ is an arbitrary
frequency. If we define
\begin{equation}
\mathcal{E}=\frac{E}{m\omega _{0}^{2}},\;U(x)=\frac{V(x)}{m\omega _{0}^{2}}
\end{equation}
then we obtain the equations of motion of a particle of unit mass; for
example:
\begin{equation}
\mathcal{E}=\frac{\dot{x}^{2}}{2}+U(x)  \label{eq:dim_energy}
\end{equation}
and
\begin{equation}
T=\frac{\sqrt{2}}{\omega _{0}}\int_{x_{-}}^{x_{+}}\frac{dx}{\sqrt{\mathcal{E}%
-U(x)}}.  \label{eq:T_U}
\end{equation}

It is worth noticing that equation (\ref{eq:dim_energy}) is not
dimensionless because $\mathcal{E}$ and $U(x)$ have units of length squared.
In order to get a truly dimensionless equation we should define a
dimensionless coordinate $q=x/L$, where $L$ has units of length. Thus $%
E/\left( m\omega _{0}^{2}L^{2}\right) $ and $V(Lq)/\left( m\omega
_{0}^{2}L^{2}\right) $ are the dimensionless counterparts of the total and
potential energies, respectively. In this paper, however, we have opted for
equations that are similar to those often found in current literature.

For example, if
\begin{equation}
V(x)=\frac{v_{2}x^{2}}{2}+\frac{v_{4}x^{4}}{4}  \label{eq:Vx2x4}
\end{equation}
we may choose $\omega _{0}=\sqrt{v_{2}/m}$ so that
\begin{equation}
U(x)=\frac{x^{2}}{2}+\frac{\lambda x^{4}}{4},\;  \label{eq:Ux2x4}
\end{equation}
where $\lambda =v_{4}/v_{2}$.

\section{The main integral}

It follows from the discussion above that the period is proportional to an
integral of the form
\begin{equation}
I=\int_{x_{-}}^{x_{+}}\frac{dx}{\sqrt{Q(x)}},  \label{eq:I_Q}
\end{equation}
were $Q(x)$ exhibits simple zeros at $x_{-}$ and $x_{+}$ and is positive
definite for all $x_{-}<x<x_{+}$. That is to say, we can write
\begin{equation}
Q(x)=(x_{+}-x)(x-x_{-})R(x)  \label{eq:Q_R}
\end{equation}
where $R(x)>0$ for all $x_{-}\leq x\leq x_{+}$.

The reason for rewriting our problem in this somewhat abstract way is that
the integral (\ref{eq:I_Q}) applies to problems other than the period of a
motion in one dimension. We will mention some examples later on.

In order to develop suitable exact and approximate expressions for the
integral (\ref{eq:I_Q}) we define the reference function
\begin{equation}
Q_{0}(x)=\frac{\omega ^{2}}{2}\left( x_{+}-x\right) \left( x-x_{-}\right)
\label{eq:Q0}
\end{equation}
that satisfies the appropriate boundary conditions at the turning points. It
is clear that $Q_{0}(x)$ is the function that would appear in the treatment
of a harmonic oscillator. Then we rewrite (\ref{eq:I_Q}) as
\begin{equation}
I=\int_{x_{-}}^{x_{+}}\ \frac{dx}{\sqrt{Q_{0}(x)}\sqrt{1+\Delta (x)}}\ ,
\label{eq:I_Delta}
\end{equation}
where
\begin{equation}
\Delta (x)\equiv \frac{Q(x)-Q_{0}(x)}{Q_{0}(x)}=\frac{2R(x)-\omega ^{2}}{%
\omega ^{2}}.  \label{eq:Delta(x)}
\end{equation}
The change of variables
\begin{equation}
x=\frac{x_{+}+x_{-}}{2}+\frac{x_{+}-x_{-}}{2}\ \cos \theta
\label{eq:x(theta)}
\end{equation}
makes the integral (\ref{eq:I_Delta}) much simpler:
\begin{equation}
I=\frac{\sqrt{2}}{\omega }\ \int_{0}^{\pi }\ \frac{d\theta }{\sqrt{1+\Delta }%
}.  \label{eq:I_theta}
\end{equation}

This equation leads to an exact expression for the period, which in most
cases one has to calculate numerically. In order to derive simple analytical
formulas we expand
\begin{equation}
\frac{1}{\sqrt{1+\Delta }}=\sum_{j=0}^{\infty }\left(
\begin{array}{c}
-1/2 \\
j
\end{array}
\right) \Delta ^{j}
\end{equation}
where $\left(
\begin{array}{l}
a \\
b
\end{array}
\right) =a!/[b!(a-b)!]$ is a combinatorial number. Notice that this series
converges for all $x$ such that $|\Delta |<1$. We thus obtain a series for
the integral (\ref{eq:I_theta}):
\begin{equation}
I=\sum_{j=0}^{\infty }I_{j},\;I_{j}=\frac{\sqrt{2}}{\omega }\ \left(
\begin{array}{c}
-1/2 \\
j
\end{array}
\right) \int_{0}^{\pi }\Delta ^{j}\,d\theta .  \label{eq:I_series}
\end{equation}
In this way we can derive approximate expressions for the integral (\ref
{eq:I_theta}) by means of the partial sums:
\begin{equation}
I^{(N)}=\sum_{j=0}^{N}I_{j}.  \label{eq:I^(N)}
\end{equation}

\section{The Duffing oscillator}

The potential--energy function (\ref{eq:Ux2x4}) gives rise to the Duffing
oscillator. Since it is parity invariant ($U(-x)=U(x)$) then $x_{+}=-x_{-}=A$
is the amplitude of the oscillations. According to the general discussion of
the preceding section, it follows from
\begin{equation}
Q(x)=\mathcal{E}-U(x)=\left( A^{2}-x^{2}\right) \left[ \frac{1}{2}+\frac{%
\lambda }{4}\left( A^{2}+x^{2}\right) \right]
\end{equation}
that
\begin{equation}
R(x)=\frac{1}{2}+\frac{\lambda }{4}\left( A^{2}+x^{2}\right)
\label{eq:R_Duffing}
\end{equation}
and
\begin{equation}
\Delta =\frac{1+\lambda A^{2}-\omega ^{2}-\frac{\lambda A^{2}}{2}\sin
^{2}\theta }{\omega ^{2}}  \label{eq:Delta(theta)_Duffing}
\end{equation}
where we have substituted $x=A\cos \theta $. We conclude that the period
depends on the dimensionless parameter $\rho =\lambda A^{2}$ that is the
ratio of $v_{4}A^{4}$ and $v_{2}A^{2}$ both having units of energy.

If we choose $\omega =\sqrt{1+\rho }$ then we obtain an already known
suitable compact expression for the integral \cite{N81}
\begin{equation}
I=\frac{\sqrt{2}}{\sqrt{1+\rho }}\int_{0}^{\pi }\frac{d\,\theta }{\sqrt{%
1-\xi \sin ^{2}\theta }},\;\xi =\frac{\rho }{2\rho +2}.
\label{eq:I_Duffing_1}
\end{equation}
This equation yields the series
\begin{equation}
I=\frac{\sqrt{2}\pi }{\sqrt{1+\rho }}\sum_{j=0}^{\infty }\left(
\begin{array}{c}
-1/2 \\
j
\end{array}
\right) ^{2}\xi ^{j}  \label{eq:I_Duff_series_1}
\end{equation}
that converges for all $|\xi |<1$; that is to say, for all $\rho >-2/3$ or $%
\rho <-2$.

When $\lambda <0$ the potential exhibits two barriers of height $%
1/(-4\lambda )$ at $x=\pm 1/\sqrt{-\lambda }$ and therefore the amplitude of
the periodic motion cannot be greater than $A_{L}=1/\sqrt{-\lambda }$. In
other words, there is periodic motion if $\rho >\rho _{L}=\lambda
A_{L}^{2}=-1$. The series (\ref{eq:I_Duff_series_1}) does not converge for $%
-1<\rho <-2/3$ and the analytical expressions that we may derive from it
will not be valid for all the values of the energy that give rise to
periodic motion. Can we improve this approach?. The answer is ''yes '' as we
will see below.

Let $R_{M}$ and $R_{m}$ be the maximum and minimum values of $R(x)$ in the
interval $[x_{-},x_{+}]$ and $\Delta _{M}$ and $\Delta _{m}$ the
corresponding values of $\Delta (x)$. Since $R(x)$ is positive definite we
know that $R_{M}\geq R(x)\geq R_{m}>0$. If we choose the value of the
adjustable parameter $\omega $ so that $\Delta _{M}=-\Delta _{m}$ we obtain
\begin{equation}
\omega _{b}^{2}=R_{M}+R_{m}>0  \label{eq:omega_b}
\end{equation}
and
\begin{equation}
\Delta _{b}(x)=\frac{2R(x)-R_{M}-R_{m}}{R_{M}+R_{m}}.  \label{eq:Delta_b}
\end{equation}
The subscript $b$ indicates that this particular value of $\omega $
''balances'' the maximum and minimum values of $\Delta (x)$. Notice that $%
|\Delta (x)|<1$ for all $x_{-}\leq x\leq x_{+}$ because $\Delta _{M}=\left(
R_{M}-R_{m}\right) /\left( R_{M}+R_{m}\right) <1$.

For the particular case of the Duffing oscillator we have $%
R_{m}=R(0)=1/2+\rho /4$ and $R_{M}=R(\pm A)=1/2+\rho /2$ so that
\begin{equation}
\omega _{b}^{2}=\frac{4+3\rho }{4}  \label{eq:omega_b_Duff}
\end{equation}
and
\begin{equation}
\Delta _{b}(\theta )=\frac{\rho }{4+3\rho }\cos (2\theta ).
\label{eq:Delta_b_Duff}
\end{equation}
Thus the integral becomes
\begin{equation}
I=\frac{2\sqrt{2}}{\sqrt{4+3\rho }}\int_{0}^{\pi }\frac{d\theta }{\sqrt{%
1+\xi \cos (2\theta )}},\;\xi =\frac{\rho }{4+3\rho }  \label{eq:I_Duffing_2}
\end{equation}
that gives rise to the series
\begin{equation}
I=\frac{2\sqrt{2}\pi }{\sqrt{4+3\rho }}\sum_{j=0}^{\infty }(-1)^{j}\left(
\begin{array}{c}
-1/2 \\
j
\end{array}
\right) \left(
\begin{array}{c}
-1/2 \\
2j
\end{array}
\right) \xi ^{2j}.  \label{eq:I_Duff_series_2}
\end{equation}
which converges for all $|\xi |<1$; that is to say, for all $\rho >-1$ or $%
\rho <-2$. In this way we may obtain simple analytical expressions for the
period valid for all values of the energy consistent with periodic motion.

According to equation (\ref{eq:T_U}) the period is given by
\begin{equation}
T=\frac{\sqrt{2}}{\omega _{0}}I  \label{eq:T_I}
\end{equation}
and equation (\ref{eq:I_Duff_series_2}) enables us to derive simple
analytical approximate expressions for it. For concreteness and simplicity
we choose $\omega _{0}=1$ in what follows. For example, the first two
approximations are
\begin{equation}
T^{(0)}=\frac{4\pi }{\sqrt{4+3\rho }}
\end{equation}
and
\begin{equation}
T^{(1)}=\frac{\pi \left( 147\rho ^{2}+384\rho +256\right) }{4\left( 4+3\rho
\right) ^{5/2}}.
\end{equation}
These expressions are expected to be accurate for small values of $\rho $,
and in fact they give the exact result for $\rho =0$. However, they are also
accurate for extremely great values of $\rho $. Notice that
\begin{equation}
\lim_{\rho \rightarrow \infty }\sqrt{\rho }T=4\int_{0}^{\pi }\frac{d\theta }{%
\sqrt{3+\cos (2\theta )}}\approx 7.4162987
\end{equation}
A straightforward calculation shows that
\begin{subequations}
\begin{eqnarray}
\lim_{\rho \rightarrow \infty }\sqrt{\rho }T^{(0)} &=&\frac{4\pi }{\sqrt{3}}%
\approx 7.26  \\
\lim_{\rho \rightarrow \infty }\sqrt{\rho }T^{(1)} &=&\frac{49\sqrt{3}\pi }{%
36}\approx 7.406.
\end{eqnarray}
\end{subequations}
We conclude that such simple analytical expressions are sufficiently
accurate for most purposes and that one easily improves them by
straightforward addition of more terms of the series (\ref
{eq:I_Duff_series_2}).

\section{Quadratic--cubic oscillator}

Parity--invariant oscillators exhibit symmetric turning points; if the
potential is nonsymmetric so are the turning points. The simplest example is
\begin{equation}
V(x)=\frac{v_{2}}{2}x^{2}+\frac{v_{3}}{3}x^{3}.  \label{eq:Vx2x3}
\end{equation}
If we again choose $\omega _{0}=\sqrt{v_{2}/m}$ then we obtain
\begin{equation}
U(x)=\frac{x^{2}}{2}+\frac{\lambda }{3}x^{3},\;\lambda =\frac{v_{3}}{v_{2}}.
\label{eq:Ux2x3}
\end{equation}
The potential--energy function $U(x)$ shows a barrier of height $%
U(x_{M})=1/\left( 6\lambda ^{2}\right) $ at $x_{M}=-1/\lambda $, and the
turning points satisfy $x_{+}>0>x_{-}$.

If we write
\begin{equation}
Q(x)=\mathcal{E}-U(x)=(x-x_{-})(x_{+}-x)(b_{0}+b_{1}x)
\end{equation}
then we obtain
\begin{equation}
b_{0}=-\frac{x_{+}x_{-}}{2(x_{+}^{2}+x_{+}x_{-}+x_{-}^{2})}\ \ \ ,\ \ \
b_{1}=\frac{\lambda }{3}=-\frac{x_{+}+x_{-}}{%
2(x_{+}^{2}+x_{+}x_{-}+x_{-}^{2})}
\end{equation}
and
\begin{equation}
\lambda =-\frac{3}{2}\ \frac{x_{-}+x_{+}}{x_{+}^{2}+x_{+}x_{-}+x_{+}^{2}}.
\end{equation}
Since $U(-x,-\lambda )=U(x,\lambda )$ we consider only the case $\lambda >0$
without loss of generality; therefore $x_{-}+x_{+}<0$ because $%
x_{+}^{2}+x_{+}x_{-}+x_{-}^{2}>0$. Taking into account that $b_{0}>0$ and $%
b_{1}>0$ we conclude that $R_{m}$ and $R_{M}$ take place at the turning
points; therefore,
\begin{equation}
\omega _{b}^{2}=R(x_{+})+R(x_{-})=-\frac{x_{+}^{2}+4x_{+}x_{-}+x_{-}^{2}}{%
2\left( x_{+}^{2}+x_{+}x_{-}+x_{-}^{2}\right) }.  \label{eq:omega_x2x3}
\end{equation}
Is $\omega _{b}$ real for all values of $\mathcal{E}$ below the barrier?. In
order to answer this question notice that the third root $x_{3}$ of $Q(x)$
is smaller than $x_{-}$ and is given by
\begin{equation}
x_{3}=-\frac{b_{0}}{b_{1}}=-\frac{x_{+}x_{-}}{x_{+}+x_{-}}<0.
\end{equation}
Therefore
\begin{eqnarray}
x_{+}^{2}+4x_{+}x_{-}+x_{-}^{2} &&  \nonumber \\
&=&\left( x_{+}+x_{-}\right) ^{2}+2x_{+}x_{-}  \nonumber \\
&=&\left( x_{+}+x_{-}\right) \left( x_{+}+x_{-}-2x_{3}\right) <0
\end{eqnarray}
because $x_{-}-x_{3}>0$ and $x_{+}-x_{3}>0$.

Finally, after the change of variables (\ref{eq:x(theta)}) the function $%
\Delta (\theta )$ takes a particularly simple form:
\begin{equation}
\Delta _{b}(\theta )=\xi \cos \theta ,\;\xi =\frac{(x_{+}^{2}-x_{-}^{2})}{%
x_{+}^{2}+4x_{+}x_{-}+x_{-}^{2}}  \label{eq:Delta_Duff_b}
\end{equation}
where $x_{+}^{2}-x_{-}^{2}<0$ because $0<x_{+}<-x_{-}$.

The resulting integral
\begin{equation}
I=\frac{\sqrt{2}}{\omega _{b}}\int_{0}^{\pi }\frac{d\theta }{\sqrt{1+\xi
\cos \theta }}  \label{eq:Ix2x3}
\end{equation}
gives rise to the series
\begin{equation}
I=\frac{\sqrt{2}\pi }{\omega _{b}}\sum_{j=0}^{\infty }(-1)^{j}\left(
\begin{array}{c}
-1/2 \\
j
\end{array}
\right) \left(
\begin{array}{c}
-1/2 \\
2j
\end{array}
\right) \xi ^{2j}.  \label{eq:Ix2x3_series}
\end{equation}
which is similar to the one derived above for the Duffing oscillator and
converges for all $|\xi |<1$.

When $\mathcal{E}=U(x_{M})$ then $x_{3}=x_{-}$ (remember that $\lambda >0$)
and $\xi =1$. We appreciate that the series (\ref{eq:Ix2x3_series})
converges for all values of the energy for which there is periodic motion.

Equation (\ref{eq:Ix2x3}) gives us a simple and exact expression for the
period of the anharmonic oscillator (\ref{eq:Ux2x3}) that requires numerical
integration to obtain results for a given set of potential parameters. On
the other hand, equation (\ref{eq:Ix2x3_series}) provides approximate
analytical expressions that one makes as accurate as desired by simply
adding a sufficiently large number of terms. The choice of one or another
depends on the particular application.

Following a different procedure Apostol \cite{A03} derived the exact
expression for the period of the quadratic--cubic oscillator
\begin{equation}
T=\sqrt{\frac{3}{2\lambda }}\frac{4}{\omega _{0}\sqrt{x_{+}-x_{3}}}%
\int_{0}^{\pi /2}\frac{d\alpha }{\sqrt{1-k^{2}\sin ^{2}\alpha }},\;k^{2}=%
\frac{x_{+}-x_{-}}{x_{+}-x_{3}}.  \label{eq:T_Apostol}
\end{equation}
The expansion of this equation in powers of $k^{2}$ is also convergent for
all values of the energy consistent with periodic motion because $k^{2}<1$.

\section{Conclusions}

In this paper we present a straightforward systematic procedure for
constructing exact and approximate expressions for the period of anharmonic
oscillators. The recipe is simple: first, we factor the function $Q(x)$ and
obtain the turning points and the function $R(x)$ as in equation (\ref
{eq:Q_R}). Second, we obtain the maximum and minimum values of $R(x)$ in the
interval between the turning points which determine the optimum value of $%
\omega $. Thus we are left with an exact expression for the period that we
may use in numerical applications. In addition to it, we may expand this
exact expression in a Taylor series in order to obtain partial sums that
become analytical expressions for the period of increasing accuracy. These
partial sums converge to the exact result for all values of the energy that
give rise to periodic motion.

The method proposed in this paper is not restricted to the period of
anharmonic oscillators with polynomial potentials. We may, for example,
expand a given arbitrary potential $U(x)$ about its minimum to any desired
degree and then apply the approach developed above. Moreover, some other
problems have been expressed in terms of integrals of the form (\ref{eq:I_Q}%
), such as, for example, the deflection of light by a massive body or the
precession of a planet orbiting around a star \cite{W72}. Recently, we have
already applied a variant of present approach to such problems \cite
{AS04,AAFS04}.

There is a wide range of interesting applications for present method and for
that reason we believe that it is suitable for teaching in advanced
undergraduate courses on classical mechanics.

\end{document}